\documentclass[aps,prb,amsmath,amssymb,superscriptaddress,reprint,
]{revtex4-2}
\usepackage{amsmath}

\usepackage{graphicx}% Include figure files
\usepackage{dcolumn}% Align table columns on decimal point
\usepackage{bm}% bold math
\usepackage{hyperref}
\hypersetup{colorlinks=true,allcolors=blue}

\usepackage{siunitx}
\usepackage{physics}
\usepackage{color}
\usepackage{mathptmx}%%%bianhei
\usepackage[normalem]{ulem}
\usepackage{braket}

\begin{document}
	%\preprint{}
	
	\title{Topological Insulator in Twisted Transition Metal Dichalcogenide Heterotrilayers} %Title of paper
	
	\author{Hao He}
	\affiliation{Hebei Provincial Key Laboratory of Photoelectric Control on Surface and Interface, School of Science, Hebei University of Science and Technology, Shijiazhuang, 050018, China}
	\author{Zhao Gong}
	\affiliation{Hebei Provincial Key Laboratory of Photoelectric Control on Surface and Interface, School of Science, Hebei University of Science and Technology, Shijiazhuang, 050018, China}
	\author{Qing-Jun Tong}
	\affiliation{School of Physical Science and Electronics, Hunan University, Changsha, 410082, People’s Republic of China}
	\author{Dawei Zhai}
	\affiliation{New Cornerstone Science Laboratory, Department of Physics, The University of Hong Kong, Hong Kong, China}
	\affiliation{HK Institute of Quantum Science \& Technology, The University of Hong Kong, Hong Kong, China}

	\author{Wang Yao}
	\affiliation{New Cornerstone Science Laboratory, Department of Physics, The University of Hong Kong, Hong Kong, China}
	\affiliation{HK Institute of Quantum Science \& Technology, The University of Hong Kong, Hong Kong, China}
	
	\author{Xing-Tao An}
	\email[Correspondence to: ]{anxt2005@163.com}
	\affiliation{Hebei Provincial Key Laboratory of Photoelectric Control on Surface and Interface, School of Science, Hebei University of Science and Technology, Shijiazhuang, 050018, China}

	\date{\today}
	
	\begin{abstract}
		The quantum spin Hall effect has been predicted in twisted homobilayer transition metal dichalcogenides (TMDs) owing to the layer-pseudospin magnetic field.
		Recently, experimental observations have also confirmed such topological states of matter.
		However, the topological electronic properties in multilayer moiré superlattices remain to be further explored.
		In twisted TMDs heterotrilayers, the realization of moiré potential with various symmetries becomes feasible.
		Here, we demonstrate that twisted trilayer TMDs can enter a topological insulator phase under the influence of moiré potential with ${C_6}$ symmetry.
		Specifically, we built two types of trilayer heterostructures, where the low-energy valence band electrons are contributed by the middle layer.
		In the AA-stacked moiré WS$_2$/WSe$_2$/MoS$_2$ heterotrilayers where only the middle layer is twisted, the maxima of the moiré potential exhibit an approximate ${C_6}$ symmetry.
		The $C_6$ symmetry effectively compensates for the spatial inversion symmetry breaking in the WSe$_2$ layer, leading to a twist-angle-dependent topological phase transition.
		Leveraging a Green's function approach, we calculate the local state density of edge states at topological minigaps, confirming their nature as moiré edge states.
		In the helical twisted AA-stacked moiré MoS$_2$/WSe$_2$/MoS$_2$ heterotrilayers, we observed a mosaic pattern of topological and trivial insulators.
		The emergence of topological mosaic is attributed to the maxima of the local moiré potential possessing $C_6$ symmetry.
		The results provide a new way for the experimental realization of topological phases in TMDs heterojunctions.
	\end{abstract}
	
	\maketitle
	
	\sloppy

	\section{\label{sec:theor}INTRODUCTION}
	
	Ever since the groundbreaking discovery of unconventional superconductivity and associated insulating states in magic-angle twisted bilayer graphene, the study of moiré physics in superlattice systems has flourished~\cite{bistritzer2011moire,cao2018unconventional,yankowitz2019tuning,cao2018correlated}. This exploration has led to the discovery of various topological phases, including the fractional quantum Hall states, in twisted bilayer graphene~\cite{tong2017topological,serlin2020intrinsic,andrews2020fractional,xie2021fractional}. The concept of moiré structures has been extended to other two-dimensional materials, for example, transition metal dichalcogenides (TMDs)~\cite{yu2017moire,wu2019topological} and magnetic materials like CrI$_{3}$~\cite{xiao2022tunable,tong2019magnetic,gong2023spin}. Recent advancements have enabled the experimental realization of trilayer moiré superlattice systems, providing alternative platforms for investigating strongly correlated physics~\cite{cao2021pauli,yang2022spectroscopy,lian2023quadrupolar}.
	
	In twisted homobilayer TMDs, intriguing Berry phase effects emerge.
	The spatially modulated interlayer tunneling and intralayer potential in twisted homobilayer TMDs endow the low-energy electrons with a layer pseudospin of the skyrmion texture. This gives rise to a real-space Berry curvature that realizes an effective Haldane model~\cite{haldane1988model} in each valley, underlying the emergence of topological moiré bands with opposite Chern numbers in the two valleys~\cite{wu2019topological,yu2020giant,zhai2020theory}.
	Experimental evidence of these topological states includes the recent discovery of the integer and fractional quantum spin Hall effect (QSHE)~\cite{Kang2024Evidence} and quantum anomalous Hall effect in twisted bilayer MoTe$_2$~\cite{cai2023signatures,zeng2023thermodynamic,Park2023Observation,Xu2023Observation}.
		
	In systems maintaining the time-reversal and spatial-inversion symmetries, the breaking of spin-inversion symmetry via spin-orbit coupling (SOC) can also induce the QSHE~\cite{kane2005quantum,kane2005z}.
	The broken spatial inversion symmetry in monolayer TMDs precludes the existence of topological states despite the strong SOC~\cite{xiao2012coupled}. However, under a spatial inversion symmetric moiré potential, the TMDs may also exhibit the QSHE.
		
\begin{figure}
		\centering
		\includegraphics[width=8.5cm]{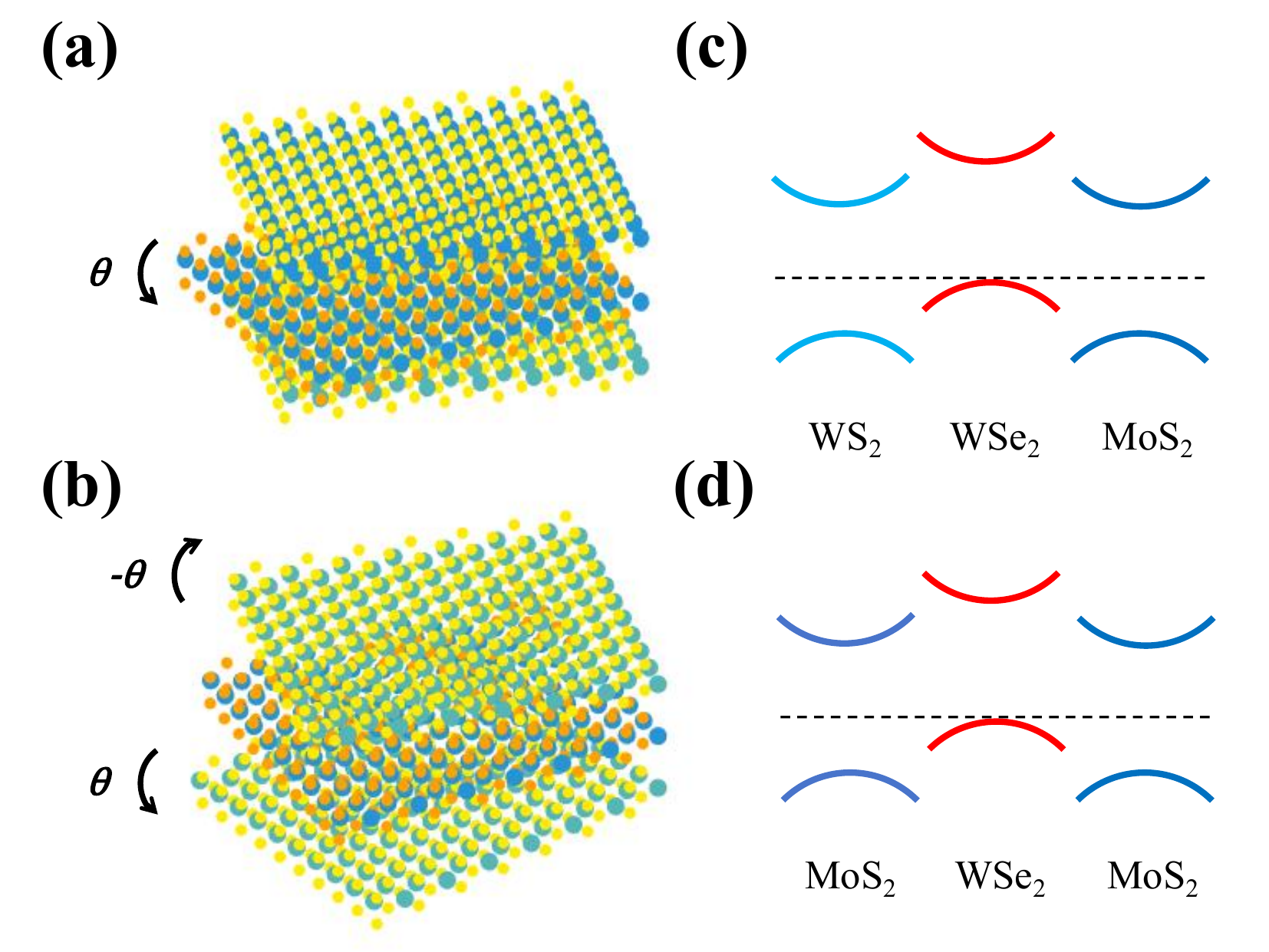}
		\caption{\label{fig:a1}(a) Schematic diagram of AA-stacked WS$_2$/WSe$_2$/MoS$_2$ heterotrilayers, where the middle layer is twisted by a small angle $\theta$ relative to the top and bottom layers. (b) Schematic diagram of AA-stacked helical MoS$_2$/WSe$_2$/MoS$_2$ heterotrilayers, where the top and bottom layers are twisted relative to the middle layer by opposite angles $\pm\theta$. (c),(d) Band alignments of WS$_2$/WSe$_2$/MoS$_2$ and MoS$_2$/WSe$_2$/MoS$_2$. The black dashed lines represent the position of the Fermi level.}
	\end{figure}

	In this paper,we built two types of heterostructures and investigated the resultant moir\'e potential to realize topological states.
	Firstly, in the AA-stacked WS$_2$/WSe$_2$/MoS$_2$ heterotrilayers, where the middle layer is twisted as illustrated in Fig.~\ref{fig:a1}(a),
	we utilized a moiré potential to characterize the impact of interlayer couplings on the valence band edge of the WSe$_2$ layer~\cite{yu2017moire,tong2020interferences}.
	The hexagonal pattern formed by the maxima of the moiré potential exhibits an approximate ${C_6}$ symmetry.
	By analyzing band structures and spin Chern numbers across various twist angles, we unveiled a topological phase transition in the system.
	The nearly $C_6$ symmetric moiré potential maxima that trap low-energy holes effectively compensates for the spatial inversion symmetry breaking of the WSe$_2$ layer, thereby transforming it into a topological insulator.
	We further constructed a finite nanoribbon system.
	Consistent with the bulk-edge correspondence, we confirmed the presence of moir\'e edge states corresponding to QSHE within the gap between the first and second moiré bands.
	Subsequently, we consider the AA-stacked  helical MoS$_2$/WSe$_2$/MoS$_2$ heterotrilayer, as illustrated in Fig.~\ref{fig:a1}(b),
	which results in a supermoiré (or moiré-of-moiré) structure with three high-symmetry regions.
	Analysis of the local band structures reveals that only one of these regions is topological.
	This is attributed to the presence of a hexagonal pattern formed by the maxima of the moiré potential in that region, which possesses $C_6$ symmetry.
	This supermoiré results in a mosaic pattern of topological insulators and trivial insulators.
	Our findings showcase the effective design of twisted TMDs heterotrilayers as topological insulators and can be generalized to realize nontrivial topology by stacking other monolayer materials exhibiting strong spin-orbit coupling yet topological triviality.
	
	\section{\label{sec:theor}MODEL AND METHODS}
	Here, we first describe the spatial structures of the AA-stacked twisted TMDs heterotrilayers. As shown in Figs.~\ref{fig:a1}(a) and (b), the middle layer is WSe$_2$ and the bottom layer is MoS$_2$, while the top layer is WS$_2$ in Fig.~\ref{fig:a1}(a) and MoS$_2$ in Fig.~\ref{fig:a1}(b). Only the middle layer is twisted by $\theta$ in Fig.~\ref{fig:a1}(a), while the three layers are twisted helically in Fig.~\ref{fig:a1}(b). This forms two sets of long-wavelength moiré patterns: the moiré-$t$ pattern between the top and middle layers, and the moiré-$b$ pattern between the bottom and middle layers.  Each moiré has three high-symmetry stacking locals within its unit cell, which are referred to as A$^{t/b}$, B$^{t/b}$, and C$^{t/b}$.
	The A$^{t/b}$ site represents the M$^{t/b}-$M$^{m}$ stacking, the B$^{t/b}$ site represents the M$^{t/b}-$X$^{m}$ stacking, and the C$^{t/b}$ site represents the X$^{t/b}-$M$^{m}$ stacking, where the metal (M) atoms or chalcogen (X) atoms of the top/bottom layer are vertically aligned with the chalcogen atoms or metal atoms of the middle layer. As shown in Figs.~\ref{fig:a1}(c) and (d), the WSe$_2$ layer and the outer layers exhibit a type-II band alignment, where the valence band edge of the heterotrilayer is contributed by WSe$_2$. Due to the significant valence band energy offset among the three TMDs layers, the electronic states near the Fermi level mainly originate from the the valence band of the WSe$_2$ layer and interlayer tunneling can be neglected. Therefore, the TMDs heterotrilayers can be approximately regarded as a WSe$_2$ monolayer with a moiré periodic potential.
	
	The moiré potential generated by each moiré can be empirically described by~\cite{tong2020interferences}
	\begin{equation}\label{eqi1}
		D(\mathbf{r}) = D_0f_0(\mathbf{r})+D_{+1}f_{+1}(\mathbf{r})+D_{-1}f_{-1}(\mathbf{r}),
	\end{equation}
	where
	\begin{equation}\label{eqi2}
		f_m(\mathbf{r}) =\frac{1}{9}|e^{-i\mathbf{K}\cdot\mathbf{r}}+e^{-i(\hat{C}_3\mathbf{K}\cdot\mathbf{r}-m\frac{2\pi}{3})} +e^{-i(\hat{C}^2_3\mathbf{K}\cdot\mathbf{r}+m\frac{2\pi}{3})}|^2.
	\end{equation}
	$\mathbf{K}$ is the wavevector at the corner of the monolayer Brillouin zone, $\mathbf{r}$ is interlayer displacement of a local stacking, which can be expressed as a function of real-space position $\mathbf{R}$. The value of $\{D_0, D_{-1}, D_{+1}\}$ can be obtained from first-principles calculations.
	The parameters for WS$_2$/WSe$_2$ are $\{32\ \mathrm{meV},20\ \mathrm{meV},120\  \mathrm{meV}\}$, and $\{65\ \mathrm{meV},81\ \mathrm{meV},183\  \mathrm{meV}\}$ for MoS$_2$/WSe$_2$.
	Finally, the moiré potential acting on the middle WSe$_2$ layer is given by
	\begin{equation}\label{eqi3}
		\mathrm{V}^\mathrm{W}(\mathbf{r}^t(\mathbf{R}),\mathbf{r}^b(\mathbf{R}))=-[D^t(\mathbf{r}^t(\mathbf{R}))+D^b(\mathbf{r}^b(\mathbf{R}))]/2.
	\end{equation}
	The absence of inversion and out-of-plane mirror symmetries in the heterotrilayers of Fig.~\ref{fig:a1} lead to a finite $\mathrm{V}^\mathrm{W}$. The local interlayer displacement $\mathbf{r}^t(\mathbf{R})$ and $\mathbf{r}^b(\mathbf{R})$ at the two moir\'e interfaces have their wavelengths and orientations separately tunable by the twisting angles $\theta$.

	In this paper, we utilize a three-band tight-binding (TB) model to describe the electronic properties around the band edge of the WSe$_2$ monolayer~\cite{liu2013Three-band}. This TB model employs three orbitals, $\{d_{z^2},d_{xy},d_{x^2-y^2}\}$, primarily contributed by the $d$ orbitals of the metal atom. The validity of this model has been demonstrated by previous studies, which accurately captures the physics near the $\pm K$ valleys, such as energy dispersion and Berry curvature. This TB model has been previously applied in various transport-related works \cite{Cortés2019Tunable,An2017Realization,Huamán2019Floquet,Natalia2020Reversible}. Under the influence of the moiré potential, the TB Hamiltonian in position space characterizing the WSe$_2$ layer reads
	\begin{equation}\label{eqi4}
		H_{TMD}= \sum_{i,\alpha} [\varepsilon_\alpha + \mathrm{V}^\mathrm{W}(\mathbf{R}_i) ] c_{i,\alpha}^\dagger c_{i,\alpha} +  \sum_{\braket{i,j},\alpha,\beta}t_{i\alpha,j\beta} c_{i,\alpha}^\dagger c_{j,\beta} +H_{SOC},
	\end{equation}
	\noindent where $\mathrm{V}^\mathrm{W}(\mathbf{R}_i)$ is the moiré potential at $\mathbf{R}_i$, $\alpha,\beta=d_{z^2},d_{xy},d_{x^2-y^2}$ are orbital indices, $c_{i,\alpha}^\dagger$ is the creation operator for electron on site $i$ with orbital $\alpha$. $t_{i\alpha,j\beta}$ is the hopping between $\alpha$ orbital at position $i$ and $\beta$ orbital at position $j$, and $\braket{,}$ denotes summation over nearest-neighbor pairs only. All the details corresponding to the hopping and on-site matrices can be found in Ref.~\cite{liu2013Three-band}.
	
	\begin{figure*}
		\centering
		\includegraphics[width=17cm]{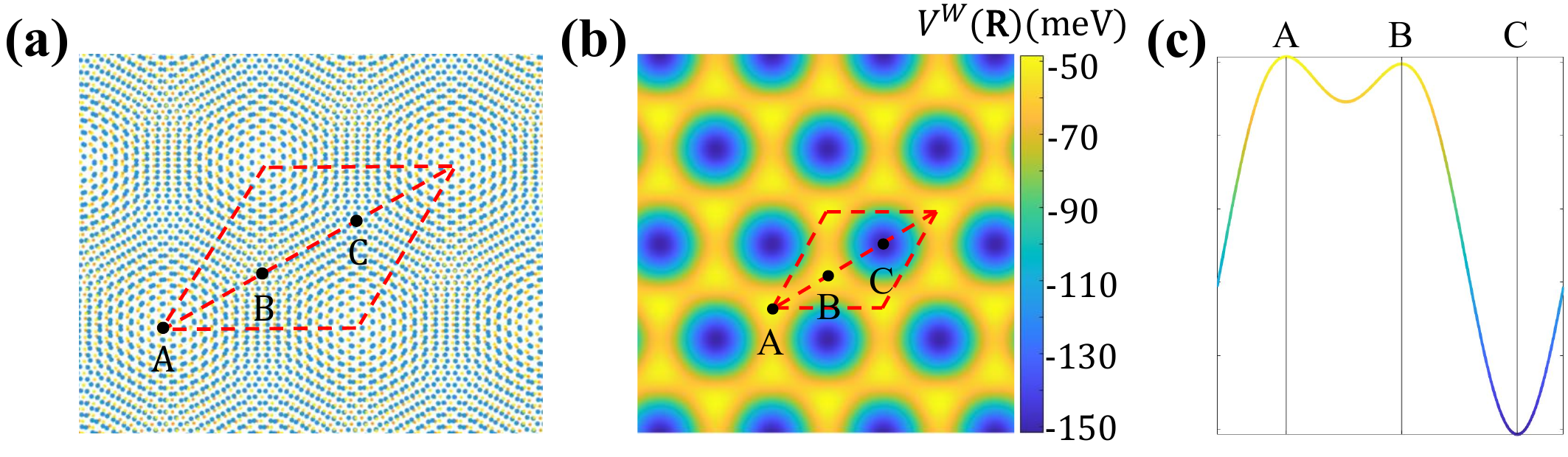}
		\caption{\label{fig:a2}(a) The moiré pattern of the twisted WS$_2$/WSe$_2$/MoS$_2$ heterotrilayers in Fig.~\ref{fig:a1}(a). The rhombus encloses a moir\'e unit cell. (b) The moiré potential pattern experienced by the middle WSe$_2$ layer in the heterotrilayers. (c) The moiré potential variation curves along the diagonal of the moiré unit cell.
		}
	\end{figure*}
	To describe the spin splitting of valence bands caused by SOC, we only consider the contribution from the on-site spin-orbit interactions of W atoms, which can be approximated as $H_{SOC}=\lambda_\mathrm{W}L_zS_z$, where $\lambda_\mathrm{W}$ is the SOC strength of the W atom, $L_z$ denotes the $z$-component of the orbital angular momentum in the basis of orbitals $\{d_{z^2},d_{xy},d_{x^2-y^2}\}$, and $S_z$ represents the $z$-component of the spin Pauli matrix. The spin splitting in the conduction band is negligible, thus the spin splitting can be represented by
	\begin{equation}\label{eqi5}
		H_{SOC}=
		\begin{bmatrix}
			0 & 0 & 0\\
			0 & 0 & i\lambda_\mathrm{W}\hat{s}_z\\
			0 & -i\lambda_\mathrm{W}\hat{s}_z & 0
			
		\end{bmatrix},
	\end{equation}
	\noindent where the SOC strength is $\lambda_\mathrm{W}=228\;\mathrm{meV}$, and $\hat{s}_z=\pm1$ represents the $z$-component of the spin degree of freedom.
	
	To explore the transport modes of the edge states, we employed a Green’s function approach to compute the local density of states (LDOS) of the charge carriers in transport~\cite{Anantram2008Reversible}. For calculating the LDOS, we utilize the formula:
	\begin{equation}\label{eqi6}
		LDOS_{L/R}(i,s,E)=\frac{\sum_{\alpha}[G^r\Gamma_{L/R}G^a]_{i\alpha,i\alpha,s}}{2\pi},
	\end{equation}
	\noindent where $L (R)$ labels the density of states induced by electron injection from the left (right), and $s$ represents spin. Here, the Green’s function $G^r(s,E)=[G^a(s,E)]^\dagger=[EI-H_{TMD,s}-\sum_{L,R}\Sigma^r_{L/R,s}]^{-1}$ with $I$ the identity matrix and the linewidth function $\Gamma_{L/R,s}(E)=i[\Sigma^r_{L/R,s}(E)-\Sigma^a_{L/R,s}(E)]$. $H_{TMD,s}$ is the Hamiltonian of the central region for different spin-subsystems. $\Sigma^r_{L/R,s}(E)=[\Sigma^a_{L/R,s}(E)]^\dagger$ is the retarded self-energy due to the coupling to lead $L/R$. For real lead, the self-energy $\Sigma^r_{L/R,s}$ can be calculated numerically.

	\begin{figure*}
		\centering
		\includegraphics[width=17.5cm]{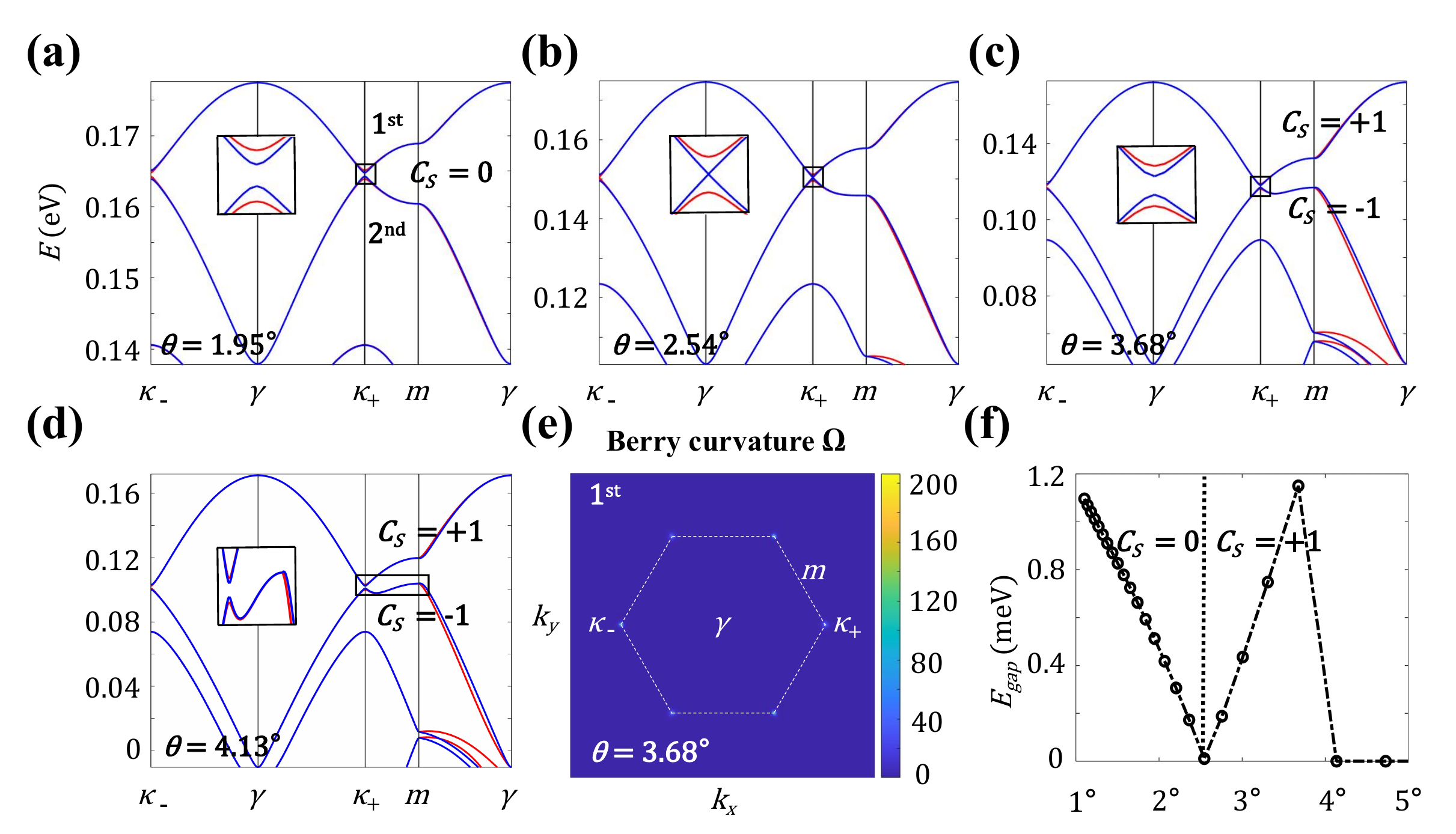}
		\caption{\label{fig:a3}(a)-(d) The band structures of twisted WS$_2$/WSe$_2$/MoS$_2$ heterotrilayers at different twist angles. At a twist angle of $\theta= 1.95^\circ$, the spin Chern numbers of the two moiré bands near the valence band maximum are zero. As the twist angle increases, the bandgap of these moiré bands closes at $\theta= 2.54^\circ$. With further increase in the twist angle, the bandgap reopens and the spin Chern numbers of the moiré bands become ($+1$, $-1$) , indicating the system transitions into a topologically nontrivial state.
			When the twist angle reaches $\theta = 4.13^\circ$, the bandgap vanishes due to the rise of the $m$ point.
			The color red (blue) represents spin-up (spin-down) states. (e) The Berry curvature of the first moiré band for spin-up. (f) Evolution of the band gap between the top two moiré bands as a function of twist angle. The system exhibits a topologically trivial phase for twist angles smaller than $\theta= 2.54^\circ$, transitioning to a topologically nontrivial phase for twist angles greater than $\theta= 2.54^\circ$.
		}
	\end{figure*}
	\section{\label{sec:theor}RESULTS AND DISCUSSION}

	\begin{figure*}
		\centering
		\includegraphics[width=17.5cm]{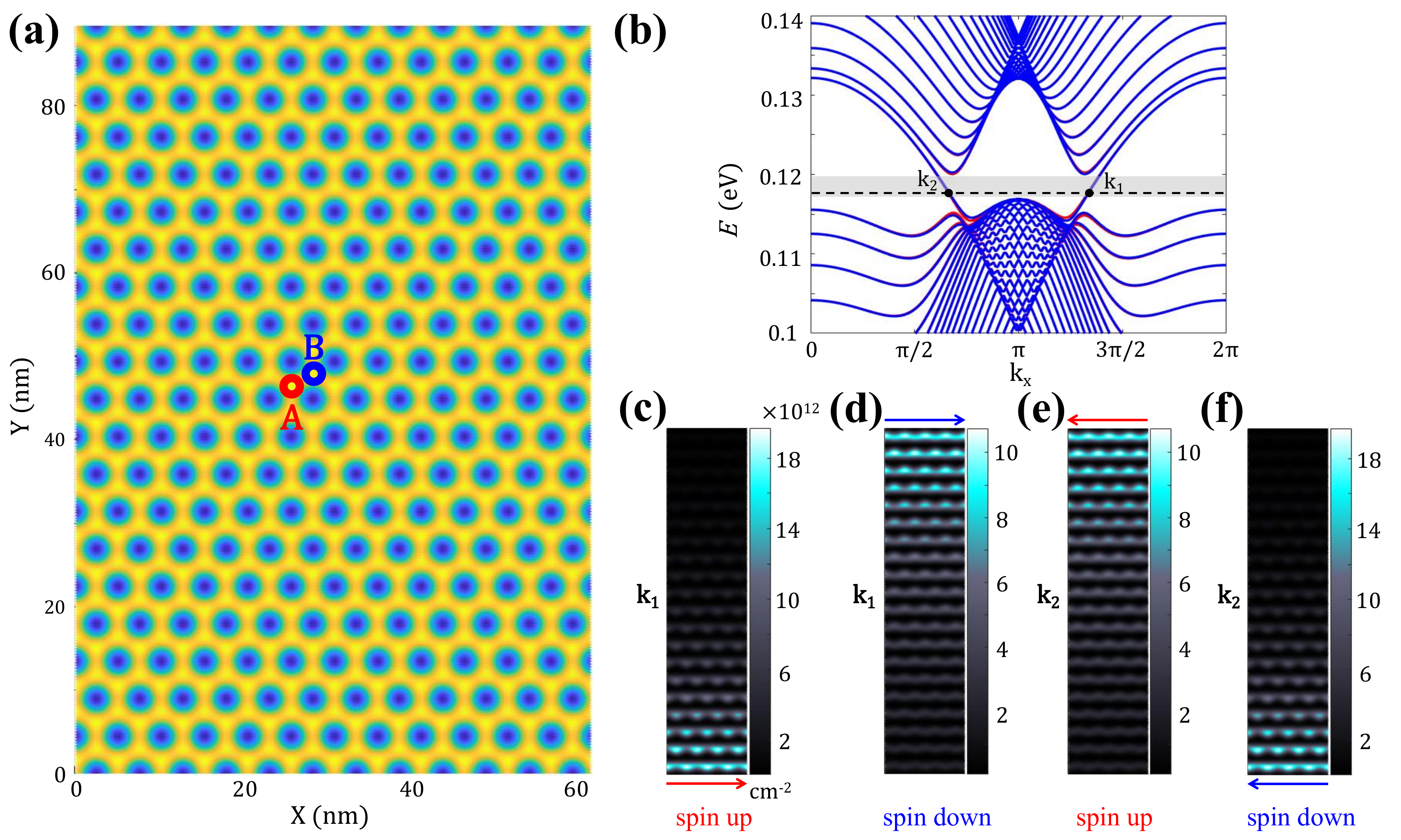}
		\caption{\label{fig:a4}(a) A finite-width WS$_2$/WSe$_2$/MoS$_2$ nanoribbon with an approximate $C_6$ symmetry moiré potential at a twist angle of $\theta= 3.68^\circ$. The high-symmetry locals A and B (cf. Fig.~\ref{fig:a2}(b)) are marked. (b) The band structure along the nanoribbon's $x$-direction at this twist angle. The shaded regions corresponds to the gap between the first and second moiré bands. (c)-(f) The local density of states in the band gap states. The arrow direction indicates the direction of the current. Different spin currents spiral along different edges, indicating a QSHE. The color red (blue) represents spin-up (spin-down) states.
		}
	\end{figure*}

	\begin{figure*}
		\centering
		\includegraphics[width=17.5cm]{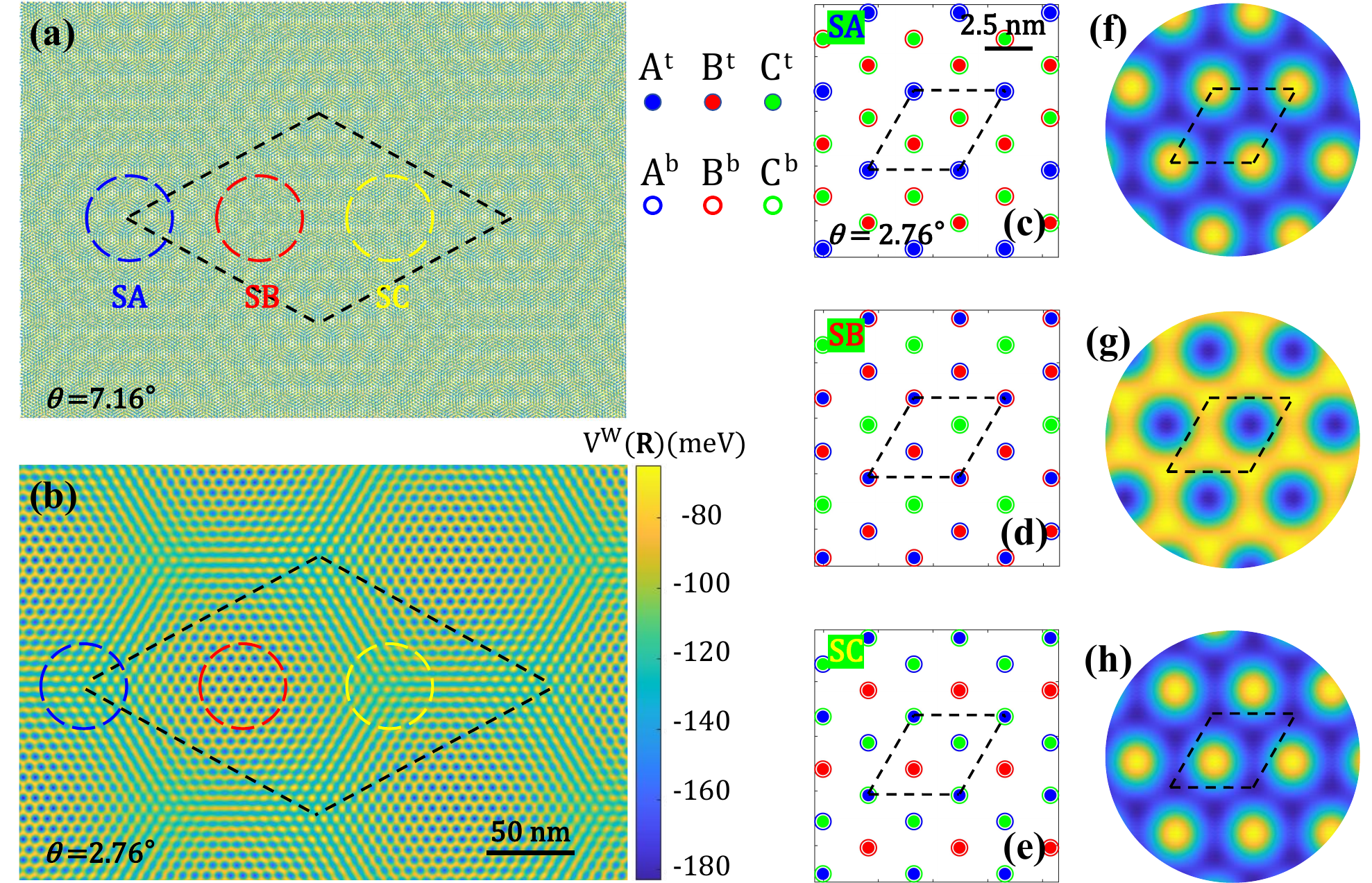}
		\caption{\label{fig:a5}(a) The supermoiré of a helical MoS$_2$/WSe$_2$/MoS$_2$ heterotrilayer in Fig.~\ref{fig:a1}(b). Three high-symmetry stacking regions are designated as SA, SB, and SC. (b) The moiré potential pattern with a rotation angle of $\theta= 2.76^\circ$.
			(c)-(e) A zoom in to the high-symmetry regions, where local high-symmetry stackings of moiré-t (moiré-b) are shown, denoted by blue, red, and green solid (empty) dots for A$^t$, B$^t$ and C$^t$ (A$^b$, B$^b$ and C$^b$), respectively. Each of these dots encloses tens to hundreds of atoms. (f)-(h) The moiré potential patterns for each high-symmetry region.
		}
	\end{figure*}
	
	\begin{figure*}
		\centering
		\includegraphics[width=17.5cm]{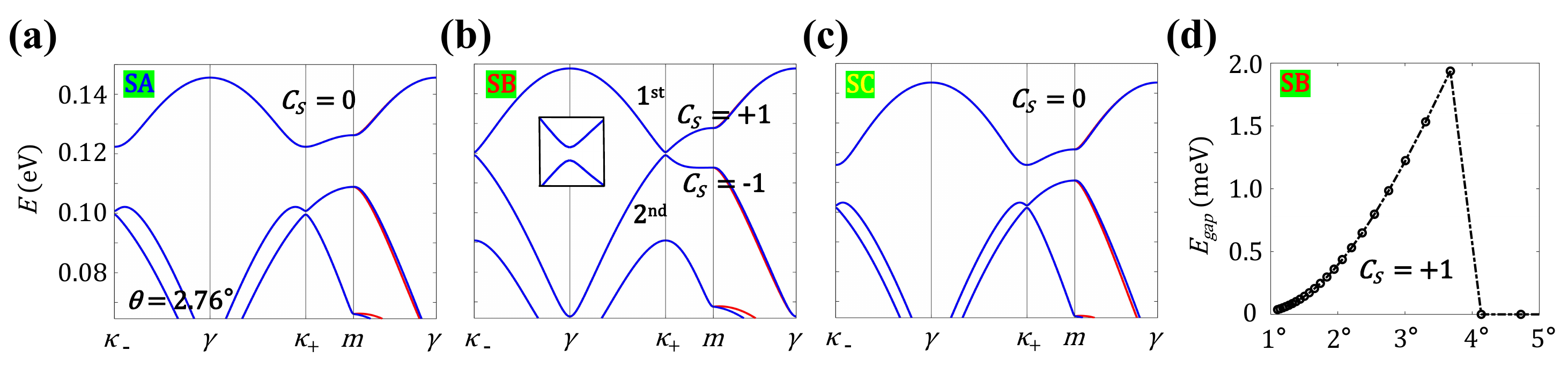}
		\caption{\label{fig:a6}(a)-(c) The band structures for the three high-symmetry regions SA, SB, and SC in a helical MoS$_2$/WSe$_2$/MoS$_2$ heterotrilayer, with a fixed twist angle of $\theta= 2.76^\circ$. Only in the SB region the band structure exhibits topological moiré bands appearing at the top of valence band. The color red (blue) represents spin-up (spin-down) states. (d) The variation of the moiré band gap in the SB region as a function of twist angle.
		}
	\end{figure*}
	
	\subsection{\label{sec:theor}WS$_2$/WSe$_2$/MoS$_2$ heterotrilayers}
	
	We first investigate the AA-stacked WS$_2$/WSe$_2$/MoS$_2$ heterotrilayers.
	Fig.~\ref{fig:a2}(a) shows the moiré superlattice, the middle layer twisted at an angle $\theta$ relative to the top and bottom layers.
	The lattice constants of the top and bottom layers are nearly identical, suggesting that the lattice constant mismatch between moiré-$t$ and moiré-$b$ is negligible.
	Furthermore, since the middle layer is twisted at the same angle relative to the top and bottom layers, this results in moiré-$t$ and moiré-$b$ having the same orientation and period. In this case, moiré-$t$ and moiré-$b$ coincide with each other.
	The moiré pattern observed here features a moiré period identical to the twisted bilayers, typically $O(10)$ nm.
	Fig.~\ref{fig:a2}(b) illustrates the moiré potential map applied to the WSe$_2$ layer in the heterotrilayers.
	In particular, the pattern formed by the maxima of the moiré potential (yellow area) closely resembles a honeycomb lattice in real space.
	Fig.~\ref{fig:a2}(c) shows the moiré potential variation along the diagonal of the moiré superlattice, encompassing the moiré potentials at the three high-symmetry stacking locals.
	It is noteworthy that, along the curve, the moiré potential energies at sites A and B are nearly identical, with values of $-48.5\;\mathrm{meV}$ and $-50.5\;\mathrm{meV}$, respectively.
	At the high-symmetry stacking locals A and B within the moiré unit cell, the energies are simultaneously elevated, favoring hole occupation when the chemical potential is near the top of the valence band.
	The hexagonal superlattice formed by the maxima of the moiré potential introduces an approximate $C_6$ symmetry to the system.
	This contrasts with the moiré potential of twisted TMDs bilayers with only $C_3$ symmetry, where energy is significantly elevated at only one of the three high-symmetry stacking locals, leading to a preference for hole occupation at a single site within the unit cell.
	The twisted TMDs heterotrilayers, featuring a moiré potential with maxima (yellow area) that form a hexagonal superlattice, may give rise to novel effects influenced by the $C_6$ symmetry.
	
	To investigate the electronic properties of the WSe$_2$ layer influenced by the moiré potential, we computed the band structure.
	Figs.~\ref{fig:a3}(a)-(d) depict the band structures of the spin-up and spin-down components at various twist angles.
	At a twist angle of $\theta= 3.68^\circ$ (Fig.~\ref{fig:a3}(c)), two moiré bands emerge at the top of valence band, each hosting two nearly degenerate subbands with differing spins.
	Moreover, a small gap appears at the $\kappa_\pm$ points, accompanied by a subtle spin splitting of the bands. This spin splitting is expected due to the non-strict $C_6$ symmetry of the moiré potential and the influence of SOC.
	Fig.~\ref{fig:a3}(e) illustrates the Berry curvature distribution of the first moiré band of spin-up at this twisting angle, indicating non-zero Berry curvatures with the same positive sign at $\kappa_\pm$ corners of the Brillouin zone.
	We computed the spin Chern numbers for the top two moiré bands by integrating the Berry curvature for each band.
	Specifically, the first band of spin-up exhibits $C_\uparrow=1$, while the second band of spin-up has $C_\uparrow=-1$.
	Due to the system's time-reversal symmetry, $C_\downarrow=-C_\uparrow$, yielding a total spin Chern number $C_s=(C_\uparrow-C_\downarrow)/2=C_\uparrow$.
	Consequently, the spin Chern number of the first (second) moiré band is $C_s=1$ ($C_s=-1$).
	The presence of non-zero spin Chern numbers in the two topmost bands suggests that when the chemical potential resides in the gap between the two moiré bands, the system behaves as a quantum spin/valley Hall insulator due to the spin valley locking in TMDs~\cite{xiao2012coupled}.
	The gap between the moiré bands varies with the twist angle, as depicted in Fig.~\ref{fig:a3}(f).
	The gap first decreases with $\theta$, it closes at $\theta= 2.54^\circ$ and reopens as $\theta$ further increases.
	This is accompanied by a topological phase transition, where the spin Chern
	number of the first band is $C_s=0$ when $\theta<2.54^\circ$, and $C_s=1$ when $\theta>2.54^\circ$.
	The distinct topological behaviors can be intuitively understood from the different
	spatial landscape of the Bloch states: at small twist angles, the kinetic energy of the holes is low (Fig.~\ref{fig:a3}(a)), the low-energy holes are more localized around the A sites of the moir\'e, forming a triangular landscape; while at large twist angles (Fig.~\ref{fig:a3}(c)), the holes are localized around both A and B sites due to larger kinetic energies, rendering a hexagonal pattern.
	
	Finally, we investigate the edge modes of WS$_2$/WSe$_2$/MoS$_2$ nanoribbons.
	At a twist angle of $\theta= 3.68^\circ$, a ribbon with the moiré potential is depicted in Fig.~\ref{fig:a4}(a).
	In the context of the hexagonal superlattice formed by the moiré potential maxima (yellow area of Fig.~\ref{fig:a2}(b)), the edges of the selected nanoribbons feature a zigzag potential configuration, while for TMDs monolayers, the edges adopt an armchair configuration.
	Fig.~\ref{fig:a4}(b) displays the energy bands of this nanoribbon system along the $x$-direction. In this scenario, the energy bands of spin-up and spin-down states are nearly degenerate. Due to the bulk-edge correspondence, topologically protected edge states with spin degeneracy appear in the gap between the first moiré band and the second moiré band as indicated by the shaded region.
	LDOS of the edge states with energy aligns with the dashed line position in Fig.~\ref{fig:a4}(b) is depicted in Figs.~\ref{fig:a4}(c)-(f).
	Right movers of spin up (down) are localized near the lower (upper) boundary of the ribbon, while their left-moving counterparts reside on the opposite sides of the ribbon.
	This implies counterclockwise (clockwise) motion for spin-up (spin-down) holes, consistent with Berry curvature $C_{\uparrow/\downarrow}=\pm1$. Analysis of edge states confirms QSHE occurrence within the bandgap region. Additionally, from Figs.~\ref{fig:a4}(c)-(f), lower edge states primarily localize at A sites of the moiré potential, while upper edge states mainly localize at B sites. These edge states, termed moiré edge states, belong to the moiré superlattice and are not localized on the nanoribbon's outermost atoms but within regions of higher moiré potential energy in these honeycomb structures.

	\subsection{\label{sec:theor}MoS$_2$/WSe$_2$/MoS$_2$ heterotrilayers}
	
	We now consider the AA-stacked MoS$_2$/WSe$_2$/MoS$_2$  helical heterotrilayers.  Fig.~\ref{fig:a5}(a) shows the moiré superlattice, the top and bottom layers are twisted with respect to the middle layer at opposite angles $\pm\theta = \pm7.16^\circ$, leading to moiré-$t$ and moiré-$b$ having the same period. In such a helically twisted system, the small misorientation between moiré-$t$ and moiré-$b$ forms a much larger supermoiré (or moiré-of-moiré) structure. Fig.~\ref{fig:a5}(b) shows the moiré potential applied to the WSe$_2$ layer under the influence of the trilayer stack, which also exhibits a supermoiré structure.
	Interestingly, within such a supermoiré structure, we observe local regions that exhibit periods analogous to the bilayer moiré.
	We identify three high-symmetry regions designated as regions SA, SB, and SC: The local points of the A$^t$, B$^t$, and C$^t$ (A$^b$, B$^b$ and C$^b$) stackings of moiré-$t$ (moiré-$b$) are represented by solid (empty) blue, red, and green dots, respectively, as shown in Figs.~\ref{fig:a5}(c)-(e).
	The moiré potentials of the three high-symmetry local structures are also shown in Figs.~\ref{fig:a5}(f)-(h).
	In particular, the maxima of the moiré potential (yellow area) in the SB region form a hexagonal pattern with $C_6$ symmetry.
	Here, the moiré potential of the A$^t-$B$^b$ stacking is equal to that of the B$^t-$A$^b$ stacking, ensuring that the $C_6$ symmetry is strict in this region.
	In contrast, in the SA and SC regions, the patterns formed by the maxima of the moiré potential exhibit only $C_3$ symmetry.
	
	The large supermoir\'e allows one to investigate the electronic structure of the WSe$_2$ layer under the influence of the moiré potential in different local regions.
	Figs.~\ref{fig:a6}(a)-(c) depict the band structures of the spin-up and spin-down components in different high-symmetry regions, with a fixed twist angle of $\theta= 2.76^\circ$.
	In the SB region's band structure (Fig.~\ref{fig:a6}(b)), two moiré bands emerge at the top of valence band, each hosting two degenerate subbands with differing spins.
	The spin Chern numbers of the top two moiré bands were found to be $+1$ and $-1$, respectively.
	This indicates that the system is a topological insulator at this region.
	Fig.~\ref{fig:a6}(d) shows the variation in the moiré band gap between different moiré bands in the SB region as a function of twist angle. The spin Chern number of the first moiré band remains consistently equal to one for all twist angles.
	In the SA and SC regions, the absence of C$_6$ symmetry in the maxima of the moiré potential results in a topologically trivial behavior.
	The supermoiré in this heterotrilayer exhibits periodic modulation of local topological order, leading to a mosaic pattern of topological and trivial insulators.

	\section{\label{sec:theor}CONCLUSIONS}

	We investigated two types of twisted TMDs heterotrilayers.
	Firstly, in the AA-stacked WS$_2$/WSe$_2$/MoS$_2$ heterotrilayers, the twisting of only the middle layer and the nearly matched lattice constants of WS$_2$ and MoS$_2$ results in the formation of a moiré potential with maxima arranged in a hexagonal pattern (exhibiting approximate C$_6$ symmetry).
	Here, we demonstrate the realization of a quantum spin/valley Hall insulator.
	The $C_6$ symmetry of the moiré potential, stemming from interlayer charge transfer, plays a pivotal role in the emergence of these topological characteristics.
	By calculating the band structure at various twist angles, we observe the emergence of two moiré bands at the top of the valence band with non-zero Chern numbers at a twist angle of $\theta= 3.68^\circ$.
	As the twist angle decreases, the bandgap progressively closes and eventually inverts, leading to a topologically trivial phase.
	Via the construction of a finite nanoribbon system, simulation of the edge states in the QSH phase has been conducted, revealing these edge states to be moiré edge states.
	Subsequently, in the AA-stacked helical MoS$_2$/WSe$_2$/MoS$_2$ heterotrilayers, a supermoiré structure with three high-symmetry regions is formed.
	In the periodically modulated supermoiré structure, we observed a mosaic pattern between topological insulators and trivial insulators.
	We calculated the band structures for the three high-symmetry regions separately and found that only in the SB region, where the moiré potential maxima feature C$_6$ symmetry, the moiré bands exhibits non-zero Chern numbers realizing quantum spin/valley Hall insulators.
	This study underscores the tunability of topological phases in van der Waals heterostructures, providing a novel approach for realizing topologically systems by engineering moiré superlattices.

	\begin{acknowledgments}
		This work was supported by the National Natural Science Foundation of China (No. 12074096) and Innovation Leading Talent Team Project in Hebei Universities. QT acknowledge support by the National Natural Science Foundation of China (No. 12374178). WY and DZ acknowledge support by the Research Grant Council of Hong Kong (AoE/P-701/20, HKU SRFS2122-7S05), and New Cornerstone Science Foundation.
	\end{acknowledgments}

	\bibliography{mymy}
	% Produces the bibliography via BibTeX.
	%\bibliographystyle{aipauth}

\end{document}